\documentclass[conference]{IEEEtran}
\IEEEoverridecommandlockouts
\usepackage{cite}
\usepackage{amsmath,amssymb,amsfonts}
\usepackage{amsthm}
\usepackage{algorithmic}
\usepackage{graphicx}
\usepackage{textcomp}
\usepackage{xcolor}
\usepackage{hyperref}
\usepackage{listings}
\usepackage{braket}
\newtheorem{theorem}{Theorem}

\newtheorem{definition}[theorem]{Definition}

\def\BibTeX{{\rm B\kern-.05em{\sc i\kern-.025em b}\kern-.08em
    T\kern-.1667em\lower.7ex\hbox{E}\kern-.125emX}}

\renewcommand{\figureautorefname}{Figure~\negthinspace}
\renewcommand{\equationautorefname}{Equation~\negthinspace}
\renewcommand{\tableautorefname}{Table~\negthinspace}

\begin{document}

\title{Evolutionary Optimization for Designing Variational Quantum Circuits with High Model Capacity

\thanks{The views expressed in this article are those of the authors and do not represent the views of Wells Fargo. This article is for informational purposes only. Nothing contained in this article should be construed as investment advice. Wells Fargo makes no express or implied warranties and expressly disclaims all legal, tax, and accounting implications related to this article.}
}

\author{\IEEEauthorblockN{ Samuel Yen-Chi Chen}
\IEEEauthorblockA{
\textit{Wells Fargo}\\
New York, NY, USA \\
yen-chi.chen@wellsfargo.com}
}

\maketitle

\begin{abstract}
Recent advancements in quantum computing (QC) and machine learning (ML) have garnered significant attention, leading to substantial efforts toward the development of quantum machine learning (QML) algorithms to address a variety of complex challenges.
The design of high-performance QML models, however, requires expert-level knowledge, posing a significant barrier to the widespread adoption of QML. Key challenges include the design of data encoding mechanisms and parameterized quantum circuits, both of which critically impact the generalization capabilities of QML models.
We propose a novel method that encodes quantum circuit architecture information to enable the evolution of quantum circuit designs. In this approach, the fitness function is based on the effective dimension, allowing for the optimization of quantum circuits towards higher model capacity.
Through numerical simulations, we demonstrate that the proposed method is capable of discovering variational quantum circuit architectures that offer improved learning capabilities, thereby enhancing the overall performance of QML models for complex tasks.
\end{abstract}

\begin{IEEEkeywords}
Quantum Machine Learning, Quantum Neural Networks, Variational Quantum Circuits, Quantum Architecture Search, Evolutionary Algorithms
\end{IEEEkeywords}

\section{Introduction}
Quantum computing (QC) holds the potential to achieve substantial speed-ups over classical computing in solving various computationally hard problems \cite{nielsen2010quantum}. At the same time, advancements in artificial intelligence and machine learning (AI/ML) technologies are enabling computational agents to achieve human-level or even superhuman performance across a range of tasks. A natural question arises: how can these two technologies be integrated to harness quantum advantages?
Despite the limitations of current quantum devices, such as imperfections and a constrained number of qubits, a hybrid computing framework that combines both quantum and classical paradigms has been proposed. Variational Quantum Algorithms (VQAs) \cite{bharti2022noisy} represent a class of hybrid algorithms where quantum parameters are optimized using classical techniques, such as gradient-based optimizers or gradient-free meta-heuristics. VQAs enable the development of architectures like Quantum Neural Networks (QNNs) to tackle a range of AI/ML tasks, including classification \cite{chen2021end, qi2023qtnvqc, mitarai2018quantum, chen2022quantumCNN}, time-series prediction \cite{chen2022quantumLSTM}, natural language processing \cite{li2023pqlm, yang2022bert, di2022dawn, stein2023applying}, and reinforcement learning \cite{chen2022variationalQRL,chen2020QRL,chen2024efficient,meyer2022survey,chen2023quantumLSTM_RL,skolik2021quantum,lockwood2020reinforcement,jerbi2021variational,CHEN2023321Async,coelho2024vqc}.
The design of high-performance Quantum Neural Network (QNN) architectures requires expertise in quantum information science to develop models that can efficiently encode input signals into quantum states and employ sequences of quantum gates that facilitate an effective learning process.
In this paper, we address this challenge using Evolutionary Quantum Architecture Search (EvoQAS) to discover circuits with a high effective dimension. Specifically, we propose a quantum circuit architecture encoding method to represent QNN models, enabling a mutation mechanism to modify this representation. After several generations of evolution, the resulting circuit representation is used to generate QNN models with a high effective dimension, which corresponds to increased model capacity.
\section{Related Work}
Quantum Architecture Search (QAS) seeks to discover efficient and high-performance quantum circuits for tasks such as generating desired quantum states \cite{kuo2021quantum, ye2021quantum, sunkel2023ga4qco, zhu2023quantum, chen2023QRL_QAS, selig2023deepqprep, sun2024quantum}, identifying optimal circuits for solving chemical ground states \cite{ostaszewski2021reinforcement, sun2024quantum}, addressing optimization problems \cite{yao2022monte, sun2024quantum}, or conducting machine learning tasks \cite{dai2024quantum, ding2022evolutionary, zhang2023evolutionary, ding2023multi,subasi2023toward,sun2023differentiable,zhang2021neural,du2022quantum}. Techniques such as reinforcement learning (RL) \cite{kuo2021quantum, ye2021quantum, dai2024quantum} and evolutionary algorithms \cite{ding2022evolutionary, zhang2023evolutionary, ding2023multi, sunkel2023ga4qco} have been applied to perform QAS. This paper differs from previous work in that we are not focused on finding a quantum circuit architecture for a specific ML task or for determining a chemical ground state. Instead, our search aims to identify a model that satisfies a specific metric of model complexity, which implies potential performance in areas such as model capacity and generalization capability.
\section{Quantum Neural Networks}
\begin{figure}[htbp]
\vskip -0.15in
\begin{center}
\includegraphics[width=1\columnwidth]{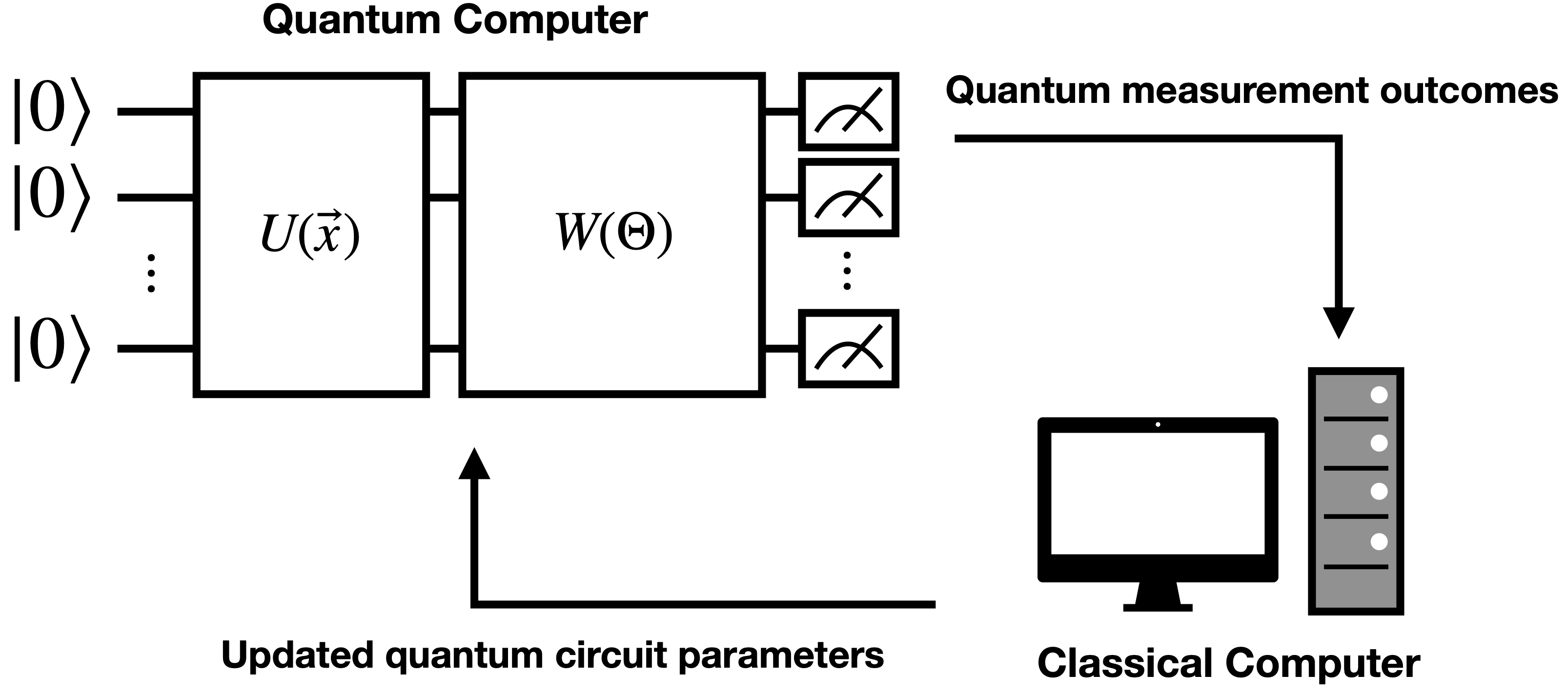}
\caption{{\bfseries Generic structure of a quantum neural network (QNN) and the scheme of hybrid quantum-classical machine learning.}}
\label{fig:generic_QNN}
\end{center}
\vskip -0.2in
\end{figure}

\begin{figure}[htbp]
\vskip -0.15in
\begin{center}
\includegraphics[width=1\columnwidth]{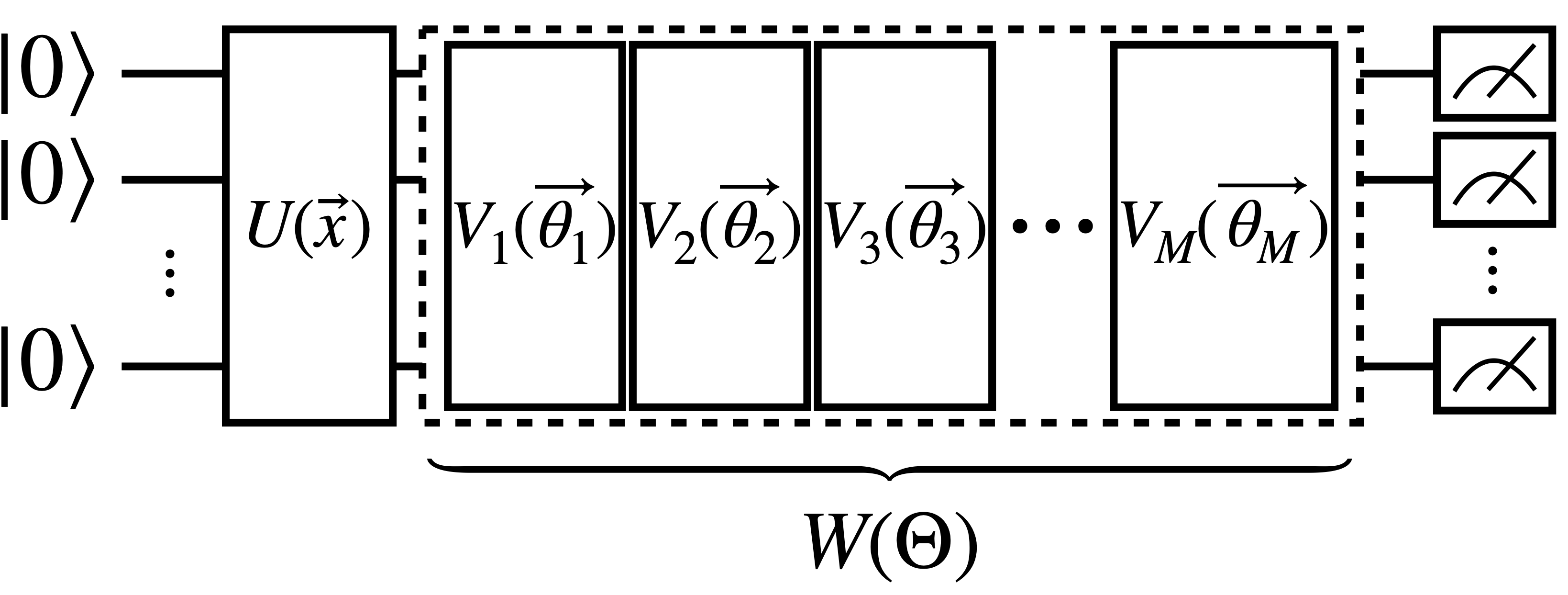}
\caption{{\bfseries Quantum neural network with multiple variational layers.}}
\label{fig:VQC_Multi_Var_Layer}
\end{center}
\vskip -0.2in
\end{figure}

The fundamental building block of a Quantum Neural Network (QNN) is the Variational Quantum Circuit (VQC), also referred to as Parameterized Quantum Circuit (PQC). The learnable parameters in VQCs are optimized using classical computing resources, as illustrated in \figureautorefname{\ref{fig:generic_QNN}}. A VQC typically consists of three primary components: the \emph{encoding circuit}, the \emph{variational circuit}, and the final \emph{quantum measurement}.
The purpose of encoding circuit $U(\vec{x})$ is to transform the input vector $\vec{x}$ into a quantum state $U(\vec{x})\ket{0}^{\otimes n}$, where $\ket{0}^{\otimes n}$ is the ground state of the quantum system and $n$ represents the number of the qubit. The encoded state then go through the variational circuit and becomes $W(\Theta)U(\vec{x})\ket{0}^{\otimes n}$. Consider the scenario in which the variational (parameterized or learnable) circuit $W(\Theta)$ is constructed by multiple layers of trainable circuit layer $V_{j}(\vec{\theta_{j}})$ (depicted in \figureautorefname{\ref{fig:VQC_Multi_Var_Layer}}), denoted as $W(\Theta) = \prod_{j = M}^{1} V_{j}(\vec{\theta_{j}})$, where $\Theta$ represents the collection of all trainable parameters $\{\vec{\theta_{1}} \cdots \vec{\theta_{M}}\}$. Then the quantum state vector generated by the encoding circuit and variational circuit can be shown as, 
\begin{equation}
    \ket{\Psi} = W(\Theta) U(\vec{x})\ket{0}^{\otimes n} = \left( \prod_{j = M}^{1} V_{j}(\vec{\theta_{j}}) \right) U(\vec{x})\ket{0}^{\otimes n}
\end{equation}
Information from the VQC can be extracted by performing measurements using predefined observables, denoted as $\hat{B}_{k}$. The VQC operation can be seen as as quantum function $\overrightarrow{f(\vec{x} ; \vec{\theta})}=\left(\left\langle\hat{B}_1\right\rangle, \cdots,\left\langle\hat{B}_n\right\rangle\right)$, where $\left\langle\hat{B}_{k}\right\rangle =\left\langle 0\left|U^{\dagger}(\vec{x})W^{\dagger}(\Theta) \hat{B}_{k} W(\Theta)U(\vec{x})\right| 0\right\rangle$.
Expectation values $\left\langle\hat{B}_{k}\right\rangle$ can be estimated by conducting repeated measurements (shots) on physical quantum devices or by direct calculation when employing quantum simulation tools.
\section{Evolutionary Quantum Architecture Search}
\begin{figure}[htbp]
\vskip -0.15in
\begin{center}
\includegraphics[width=1\columnwidth]{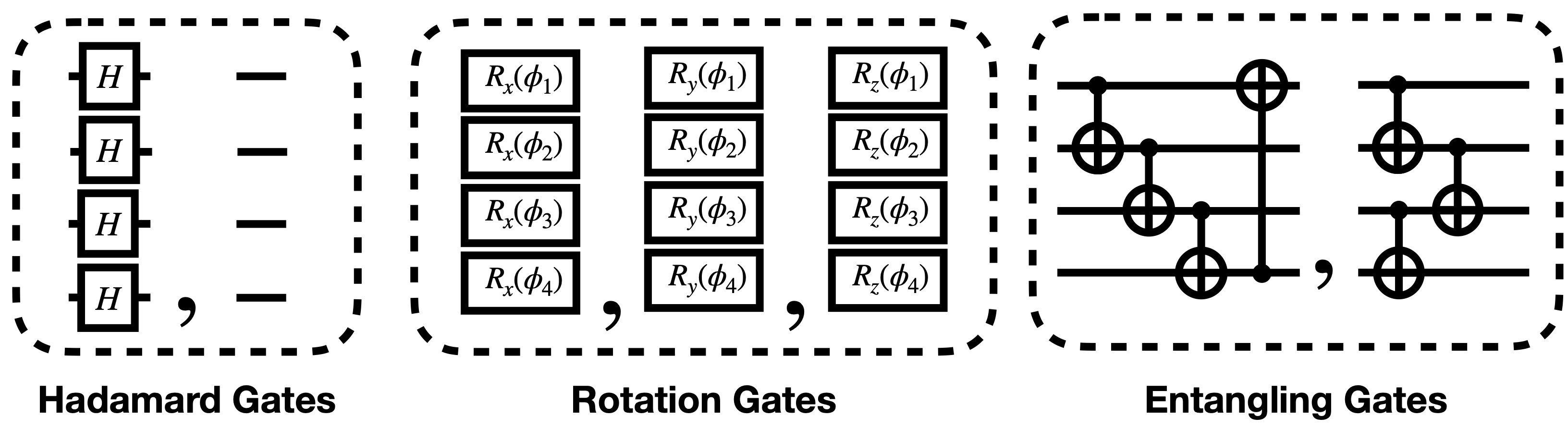}
\caption{{\bfseries Examples of allowed quantum circuit components.}}
\label{fig:allowed_ansatz}
\end{center}
\vskip -0.15in
\end{figure}

Consider the task of constructing a quantum circuit $\mathcal{C}$, which is composed of several sub-components $\mathcal{S}_{1}, \mathcal{S}_{2}, \dots, \mathcal{S}_{n}$. Each sub-component $\mathcal{S}_{i}$ is associated with a set of available circuit options $\mathcal{B}_{i}$, where $|\mathcal{B}_{i}|$ represents the number of feasible choices for that particular sub-component. Therefore, the total number of possible configurations for the circuit $\mathcal{C}$ can be expressed as $N = |\mathcal{B}_{1}| \times |\mathcal{B}_{2}| \times \dots \times |\mathcal{B}_{n}|$.
Here, our goal is to construct a quantum circuit in which the \emph{encoding sub-circuit} $\mathcal{S}_{1}$ and \emph{variational sub-circuit} $\mathcal{S}_{2}$ are to be found by certain evolutionary search algorithms. For a concrete example, we consider the case that the encoding sub-circuit can be constructed by Hadamard gates $H$ and single-qubit rotations ($R_{x}$, $R_{y}$ and $R_{z}$). The total number of possible encoding sub-circuit $\mathcal{B}_{1}$ is therefore $|\mathcal{B}_{1}| = 2 \times 3 = 6$. For the variational sub-circuit, although there can be multiple ways to entangle the qubits, let's assume that we have two possible ways to perform the entanglements and again three single-qubit rotations ($R_{x}$, $R_{y}$ and $R_{z}$) as in the encoding sub-circuit. Then, the total number of possible variational sub-circuit $\mathcal{B}_{2}$ is therefore $|\mathcal{B}_{2}| = 2 \times 3 = 6$. The total number of realization for this circuit is therefore $N = \mathcal{B}_{1} \times \mathcal{B}_{2} = 36$. The assumed circuit components in this work is shown in \figureautorefname{\ref{fig:allowed_ansatz}}.
We can further consider the circuit with multiple layers of variational sub-circuit $\mathcal{S}_{2}, \mathcal{S}_{3} \cdots, \mathcal{S}_{n}$. Circuits of this type are frequently employed when a larger number of trainable parameters is necessary to achieve the desired performance. When this is the case, the search space for the circuit grows to be $N = |\mathcal{B}_{1}| \times |\mathcal{B}_{2}| \times \cdots \times |\mathcal{B}_{n}| = |\mathcal{B}_{1}| \times |\mathcal{B}_{2}|^{n-1}$, with the assumption that there is a fixed number of allowed combinations for variational sub-circuit. 
To enable algorithmic search for quantum circuit combinations, we design the circuit representation as a dictionary data structure $\mathcal{R}$. 

\begin{align}
\mathcal{R} = \left\{
    \begin{aligned}
    &\text{“encoding\_layer”} & \mapsto \left( \mathbf{x}_1, \mathbf{x}_2 \right), \\
    &\text{“variational\_layer”} & \mapsto \left[ \left( \mathbf{y}_1^{(i)}, \mathbf{y}_2^{(i)} \right) \right]_{i=1}^{N}
    \end{aligned}
\right\}
\end{align}
where $\mathbf{x}_1 \in \mathbb{R}^{\text{NUM\_H\_LAYERS}}$, $\mathbf{x}_2 \in \mathbb{R}^{\text{NUM\_ROTATIONS}}$, $\mathbf{y}_1^{(i)} \in \mathbb{R}^{\text{NUM\_ENTANGLING}}$, $\mathbf{y}_2^{(i)} \in \mathbb{R}^{\text{NUM\_ROTATIONS}}$, $N = \text{num\_of\_var\_layers}$.
For example, in the "encoding\_layer", the vector $\mathbf{x}_1$ controls whether or not Hadamard gates $H$ are used to initialized the quantum state. Hence, the vector $\mathbf{x}_1$ is in $\mathbb{R}^2$. With the same logic, the vector $\mathbf{x}_2$ controls which kind of rotations (e.g. $R_{x}$, $R_{y}$ and $R_{z}$) is used for encoding values into quantum states, therefore $\mathbf{x}_2$ is in $\mathbb{R}^3$. 
The representation of the “variational\_layer” is consistent, where $\mathbf{y}_1^{(i)}$ governs the entangling component and $\mathbf{y}_2^{(i)}$ governs the rotation component of the $i$-th variational sub-circuit, respectively.
Given a circuit representation $\mathcal{R}$, sampling is required to get the actual circuit for performance evaluation. For example, the vector $\mathbf{x}_1 \in \mathbb{R}^2$ will be converted into probabilities by softmax function. 
The actual circuit component will be sampled from this distribution, and a corresponding one-hot representation will be generated. Once this representation is created, the circuit can be constructed based on the predefined rules, such as those outlined in
\tableautorefname{\ref{tab:one_hot_mapping}}. 
The representation $\mathcal{R}$ allows for easy mutation, as the encoding consists of floating-point numbers. By introducing small perturbations, we can subtly alter the quantum circuit properties without causing significant changes.
\begin{table}[htbp]
\vskip -0.15in
\centering
\caption{Circuit configurations and their corresponding one-hot encoding.}
\label{tab:one_hot_mapping}
\resizebox{\columnwidth}{!}{%
\begin{tabular}{|c|c|c|}
\hline
\textbf{H Layer}           & \textbf{Rotations}  & \textbf{Entangling Layers}       \\ \hline
$[1,0]$ (with $H$ gate)    & $[1,0,0]$ ($R_{x}$) & $[1,0]$ (entangling\_layer)        \\ \hline
$[0,1]$ (without $H$ gate) & $[0,1,0]$ ($R_{y}$) & $[0,1]$ (cycle\_entangling\_layer) \\ \hline
    --                     & $[0,0,1]$ ($R_{z}$) &            --                    \\ \hline
\end{tabular}%
}
\vskip -0.15in
\end{table}
\section{Methods}
In this paper, we present EvoQAS-ED, a QAS method designed to discover QNN architectures that meet a specified model complexity metric. The metric we use to assess QNN models is the \emph{effective dimension}, which provides an indication of the model's capacity \cite{abbas2021power}. Our approach adheres to the definition of effective dimension as outlined in \cite{abbas2021power}.
\begin{definition}
The \emph{effective dimension} of a statistical model $\mathcal{M}_\Theta := \{p(\cdot, \cdot; \theta) : \theta \in \Theta \}$ with respect to $\gamma \in (0,1]$, a $d$-dimensional parameter space $\Theta \subset \mathbb{R}^d$ and $n\in \mathbb{N}$, $n>1$ data samples is defined as
\begin{equation}\label{eq_dim}
d_{\gamma,n}(\mathcal{M}_{\Theta}):= 2 \frac{ \log\left( \frac{1}{V_{\Theta}} \int_{\Theta} \sqrt{\det\Big(\mathrm{id}_d + \frac{\gamma n}{2\pi \log n} \hat F(\theta) \Big) } \, \mathrm{d} \theta  \right)}{ \log\left(\frac{\gamma n}{2\pi \log n}\right)} 
\end{equation}
where $V_\Theta:=\int_{\Theta}\mathrm{d} \theta \in \mathbb{R}_+$ is the volume of the parameter space. $\hat{F}(\theta) \in \mathbb{R}^{d\times d}$ is the normalised Fisher information matrix defined as $\hat F_{ij}(\theta):= d \frac{V_{\Theta}}{\int_{\Theta} \mathrm{tr}( F(\theta)) \mathrm{d} \theta} F_{ij}(\theta)$, where the normalisation ensures that $\frac{1}{V_{\Theta}}\int_{\Theta} \mathrm{tr}( \hat F(\theta)) \mathrm{d} \theta=d$. 
\end{definition}
The \emph{Fisher information matrix} is defined as $F(\theta)
    = \mathbb{E}_{(x,y) \sim p} \Big[\frac{\partial}{\partial \theta} \log p(x,y; \theta) \frac{\partial}{\partial \theta}  \log p(x,y; \theta)^{\!\mathsf{T}} \Big] \in \mathbb{R}^{d\times d}$ and can be approximated by the \emph{empirical} Fisher information matrix $\tilde{F}_k(\theta) = \frac{1}{k} \sum_{j=1}^k \frac{\partial}{\partial \theta}  \log p(x_j,y_j; \theta) \frac{\partial}{\partial \theta} \log p(x_j,y_j; \theta)^{\!\mathsf{T}}$.
Specifically, we employ the effective dimension as the fitness function in our evolutionary search method. The primary constraint is that the QNN model must be differentiable.
\section{Experiments}
\begin{figure}[htbp]
\vskip -0.15in
\begin{center}
\includegraphics[width=1\columnwidth]{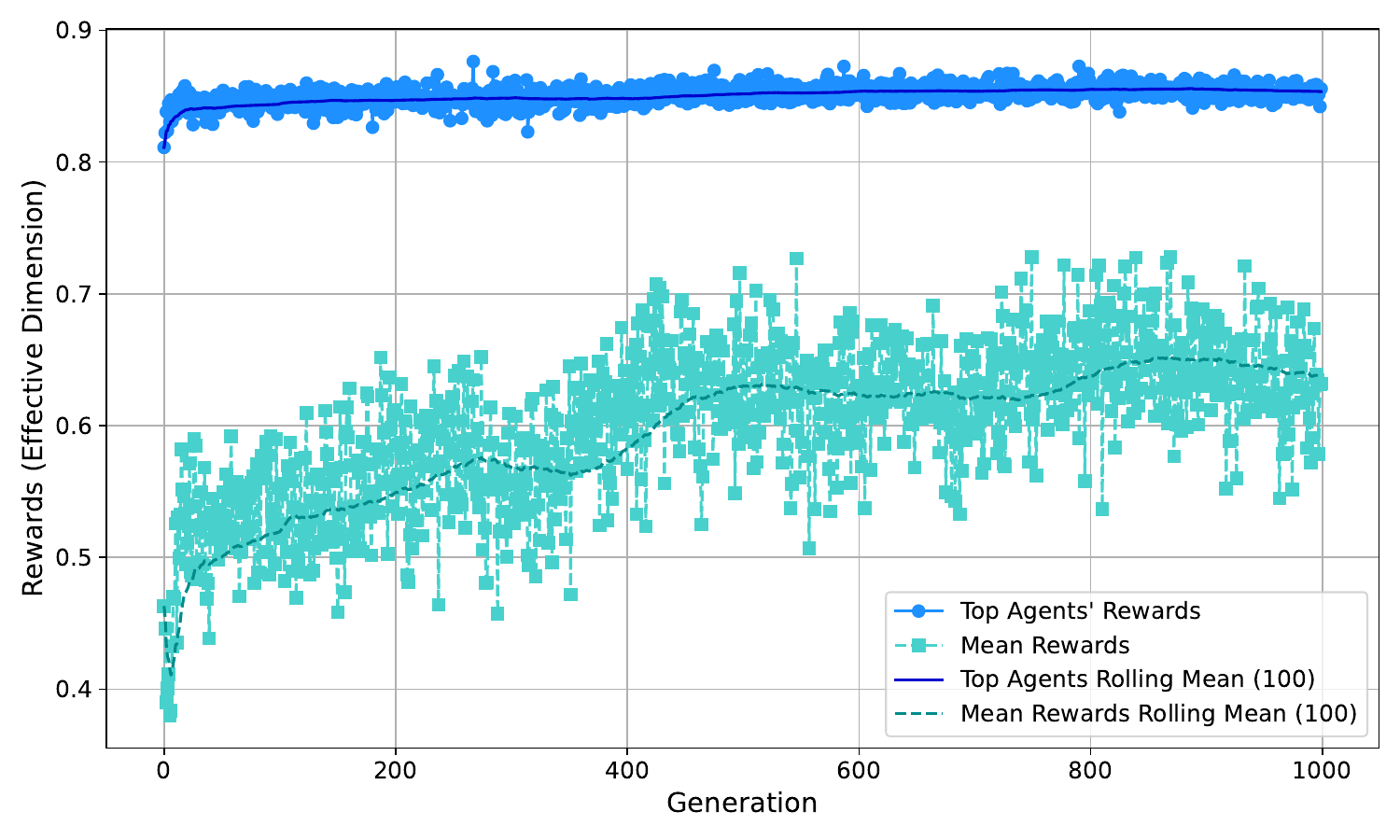}
\caption{{\bfseries Results: effective dimension (rewards) with respect to generation of evolution when data size = 1000.}}
\label{fig:ED_datasize_1000}
\end{center}
\vskip -0.15in
\end{figure}

\begin{figure}[htbp]
\vskip -0.2in
\begin{center}
\includegraphics[width=1\columnwidth]{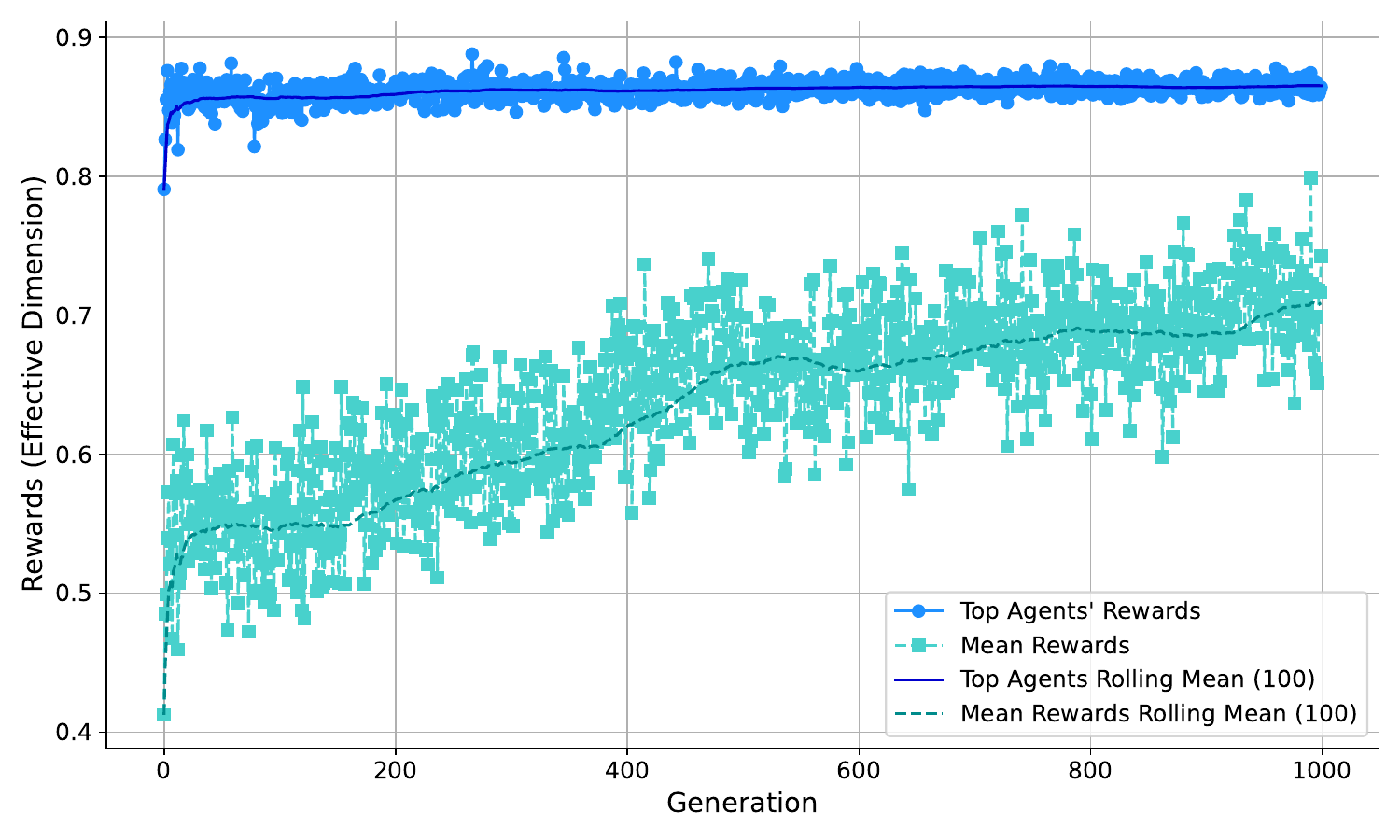}
\caption{{\bfseries Results: effective dimension (rewards) with respect to generation of evolution when data size = 2000.}}
\label{fig:ED_datasize_2000}
\end{center}
\vskip -0.2in
\end{figure}

\begin{figure}[htbp]
\vskip -0.1in
\begin{center}
\includegraphics[width=1\columnwidth]{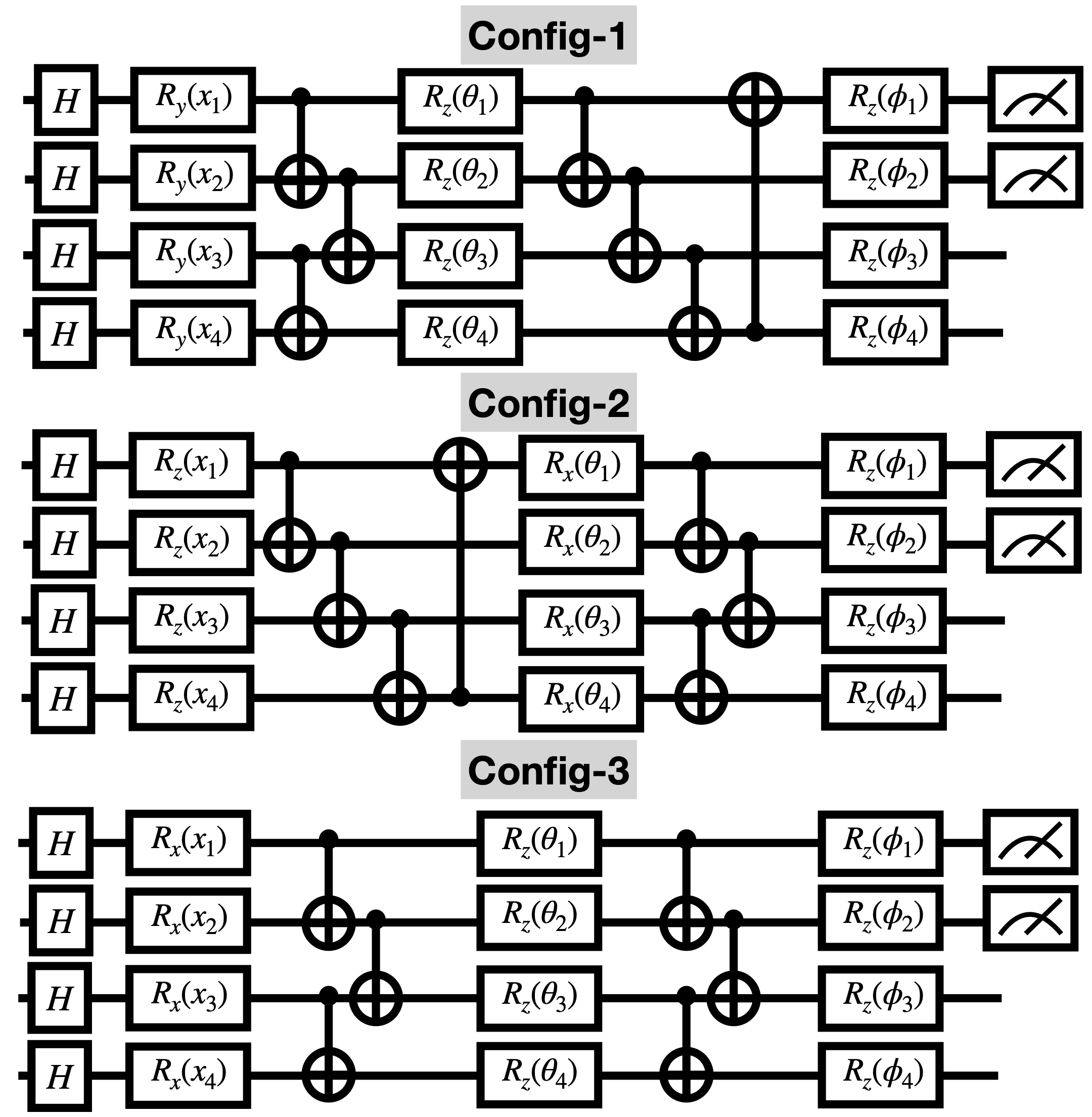}
\caption{{\bfseries Results: Example QNN architectures found by the proposed method.}}
\label{fig:example_found_circuits}
\end{center}
\vskip -0.2in
\end{figure}

\begin{figure}[htbp]
\vskip -0.15in
\begin{center}
\includegraphics[width=1\columnwidth]{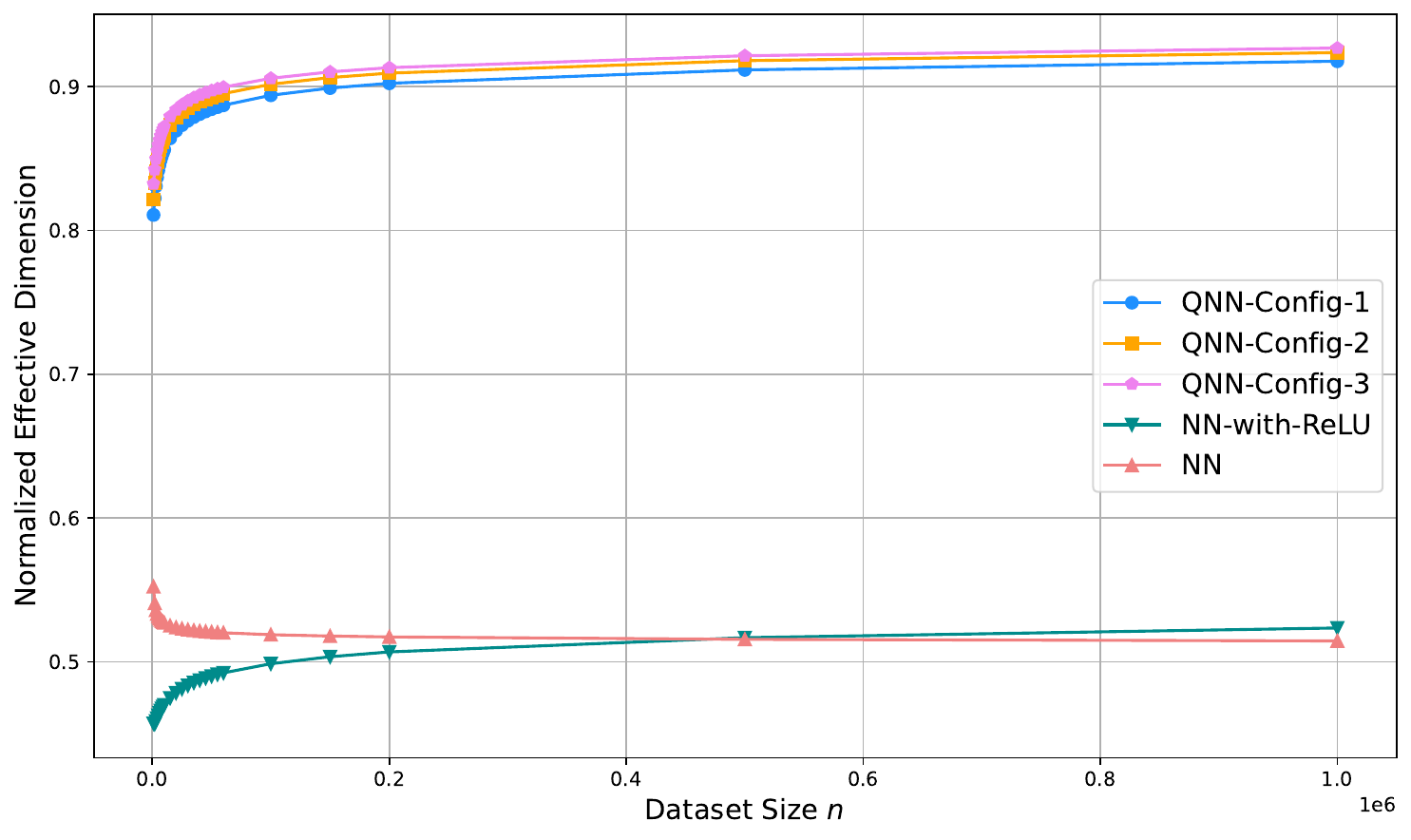}
\caption{{\bfseries Results: effective dimension with respect to different dataset size $n$ in found QNN configurations and classical NN baselines.}}
\label{fig:effective_dimension_trend_quantum_vs_classical}
\end{center}
\vskip -0.2in
\end{figure}

\begin{figure}[htbp]
\vskip -0.15in
\begin{center}
\includegraphics[width=1\columnwidth]{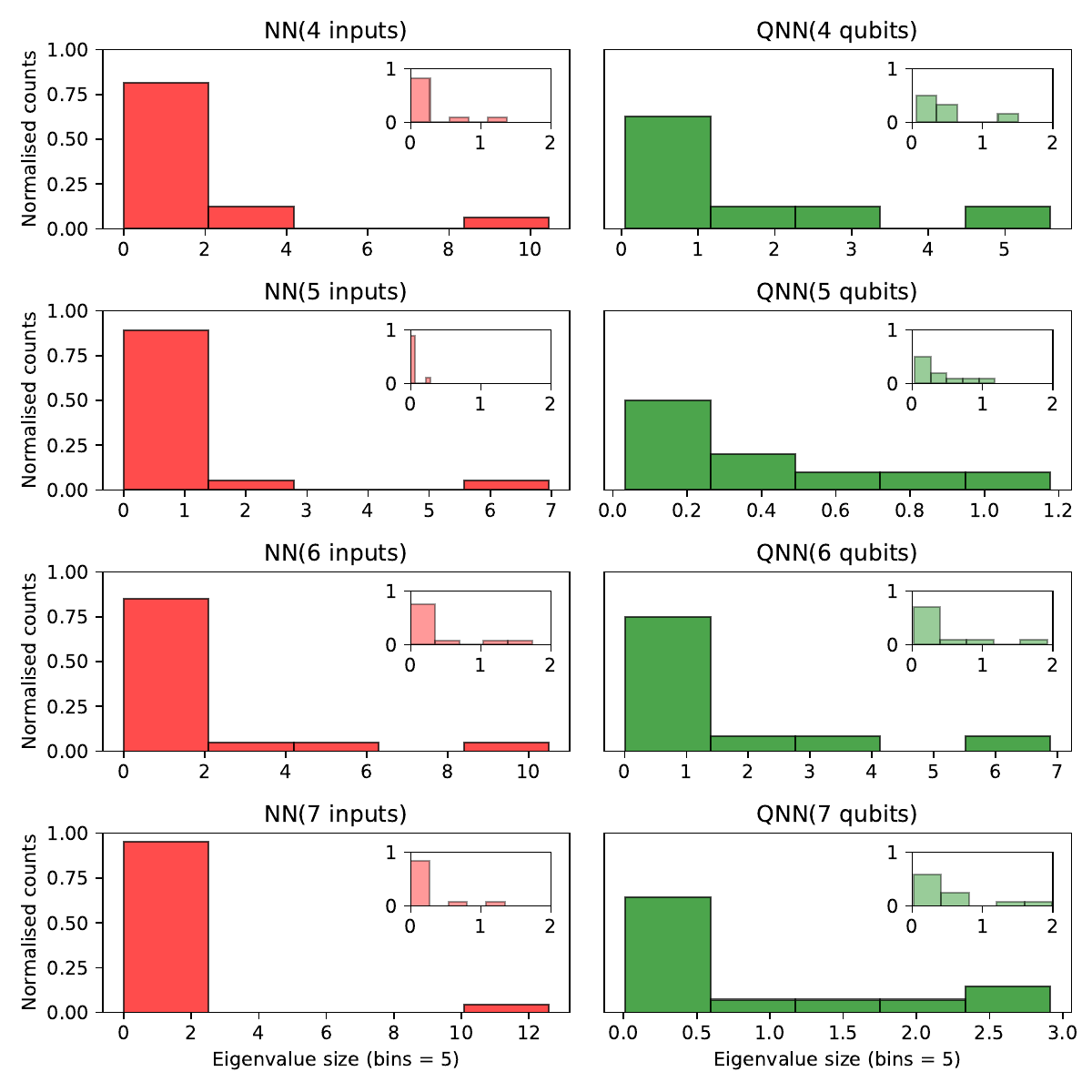}
\caption{{\bfseries Results: Comparison of eigenvalue spectrum of Fisher information matrix from QNN and NN.}}
\label{fig:compare_eigenvalue_spectrum}
\end{center}
\vskip -0.2in
\end{figure}

In EvoQAS-ED, we run simulations with a population size of $\mathcal{P} = 50$. In each generation, we select the top 10 agents with the highest effective dimension (ED) scores to serve as parents. These parents are then mutated to create new circuit representations $\mathcal{R}$ for the next generation. The mutation process consists of adding Gaussian noise to the original circuit representation, $\mathcal{R} \leftarrow \mathcal{R} + \sigma\epsilon$, where $\sigma$ (set to 0.02) represents the \emph{mutation power}, and $\epsilon$ is Gaussian noise sampled from the normal distribution $\mathcal{N}(0, I)$.
In \figureautorefname{\ref{fig:ED_datasize_1000}} and \figureautorefname{\ref{fig:ED_datasize_2000}}, we present the results of the evolutionary optimization over 1000 generations. We conducted two experiments with slightly different configurations: the dataset size $n$ used to evaluate the effective dimension $d_{\gamma, n}$. As shown in both \figureautorefname{\ref{fig:ED_datasize_1000}} and \figureautorefname{\ref{fig:ED_datasize_2000}}, there is no significant difference between the two settings. Additionally, we observe that the rewards for the top agents converge after only a few generations. These top agents are saved for further analysis.
We selected dataset sizes of $n = 1000$ and $n = 2000$ to estimate the effective dimension because current QML training often relies on smaller datasets. This preference for smaller datasets is likely due to limitations in existing quantum simulations or hardware, as well as the desire to investigate the behavior of QML models in the small data regime.
After 1000 generations of training, we sampled from the evolved quantum circuit representations with high effective dimension. A selection of circuits generated through this process is displayed in \figureautorefname{\ref{fig:example_found_circuits}}. 
We analyze the behavior of these QNN architectures using different dataset sizes $n$, as described in \equationautorefname{\ref{eq_dim}}. In most cases/models, the effective dimension $d_{\gamma,n}(\mathcal{M}_{\Theta})$ increases with $n$ and eventually saturates. However, it is not guaranteed that $d_{\gamma,n}(\mathcal{M}_{\Theta})$ will grow for all types of models, as noted in \cite{abbas2021power}.
In \figureautorefname{\ref{fig:effective_dimension_trend_quantum_vs_classical}}, we analyze the trend of effective dimension with respect to dataset size $n$. The three QNN architectures generated by the proposed evolutionary search method (as shown in \figureautorefname{\ref{fig:example_found_circuits}}) exhibit the expected behavior: the effective dimension increases and then saturates. In contrast, the two baseline classical neural networks (NNs) show different behaviors. These classical NNs are designed to have a comparable number of trainable parameters to the QNNs. The key distinction is that one of the classical NNs lacks a non-linear activation function, such as ReLU. We observe that all QNNs identified by our method achieve a significantly higher effective dimension than the classical NNs with similar model sizes. Furthermore, even without non-linear activation functions, the QNNs display an effective dimension growth trend similar to that of classical NNs equipped with non-linear activations like ReLU.
Next, we analyze the eigenvalue spectrum of the Fisher information matrix for the discovered QNNs and their corresponding classical NNs. Prior research \cite{karakida2019universal} has shown that the Fisher information matrix of non-linear classical NNs tends to be highly degenerate, with a few large eigenvalues. This leads to a flat parameter space in most directions, as indicated by near-zero eigenvalues, while certain directions are highly distorted due to a few significantly large eigenvalues. Under certain conditions, the Hessian matrix coincides with the Fisher information matrix and exhibits a similar behavior \cite{kunstner2019limitations,pennington2018spectrum,liao2018approximate}.
Such spectra have been reported to impede training and may result in suboptimal outcomes \cite{lecun2002efficient}. Similar effects have also been observed in quantum model training \cite{abbas2021power}: quantum models affected by the barren plateau phenomenon (vanishing gradient problem) exhibit Fisher information spectra with an increasing concentration of eigenvalues near zero as the number of qubits grows. On the other hand, models with Fisher information spectra not concentrated around zero are less likely to suffer from barren plateaus \cite{abbas2021power}. In this study, we numerically simulate one of our discovered QNN architectures with varying numbers of qubits to analyze their Fisher information spectra. We also compare these results to a classical NN counterpart with a comparable number of inputs (and number of parameters).
In \figureautorefname{\ref{fig:compare_eigenvalue_spectrum}}, we observe that the QNN exhibits a more evenly distributed eigenvalue spectrum of the Fisher information matrix. This distribution remains consistent as the number of qubits increases from 4 to 7. In contrast, the eigenvalues in classical NN models, with varying input sizes (and thus different model parameters), are mostly concentrated around zero. This concentration is particularly evident in the subplots of each histogram, which depict the eigenvalue distribution over smaller ranges. Additionally, unlike in the classical NN, there are no outlying eigenvalues in the QNN.
\section{Conclusion}
In this paper, we introduce EvoQAS-ED, which incorporates a quantum circuit representation method to efficiently discover QNN architectures with high effective dimension, indicating high model capacity. The proposed framework is adaptable and can be extended to various QNN metrics. Additionally, it holds the potential to integrate with existing QAS methods to search for high-performance QNN architectures tailored to a wide range of QML challenges.
\clearpage
\bibliographystyle{IEEEtran}
\bibliography{bib/qml_examples,bib/vqc,bib/fisher_info,bib/qc,bib/qas}
\end{document}